\documentclass[sigconf]{acmart}

\AtBeginDocument{%
  \providecommand\BibTeX{{%
    \normalfont B\kern-0.5em{\scshape i\kern-0.25em b}\kern-0.8em\TeX}}}

\copyrightyear{2022}
\acmYear{2022}
\setcopyright{rightsretained}
\acmConference[PDC 2022 Vol. 2]{Participatory Design Conference 2022: Volume 2}{August 19-September 1, 2022}{Newcastle upon Tyne, United Kingdom}
\acmBooktitle{Participatory Design Conference 2022: Volume 2 (PDC 2022 Vol. 2), August 19-September 1, 2022, Newcastle upon Tyne, United Kingdom}
\acmDOI{10.1145/3537797.3537799}
\acmISBN{978-1-4503-9681-3/22/08}

\begin{document}

\title[Would the Trees Dim the Lights?]{Would the Trees Dim the Lights? Adopting the Intentional Stance for More-Than-Human Participatory Design}

\author{Ned Cooper}
\email{edward.cooper@anu.edu.au}
\orcid{0000-0003-1834-279X}
\affiliation{%
  \institution{Australian National University}
  \city{Canberra}
  \country{Australia}
}

\renewcommand{\shortauthors}{}

\begin{abstract}
The 2019/20 Black Summer bushfires in Australia demonstrated the brutal and disastrous consequences of changing the technological world without considering linkages with the biophysical, ecological or human worlds. An emerging more-than-human design philosophy encourages designers to consider such interrelations between humans and non-human entities. Yet, the design research community has focused on situated or embodied experiences for designers, rather than developing processes to legitimate the perspectives of non-human entities through participatory design. This paper explores how adopting the `intentional stance', a concept from philosophy, might provide a heuristic for more-than-human participatory design. Through experimentation with the intentional stance in the context of smart lighting systems, the paper demonstrates that the approach has potential for non-human entities from the ecological world, but less so for the biophysical world. The paper concludes by encouraging critique and evolution of the intentional stance, and of other approaches, to legitimate the perspectives of non-human entities in everyday design.
\end{abstract}



\maketitle

\section{Introduction}
Climate variability and long-term climate trends increase the likelihood and severity of catastrophic weather events. These weather events can have a destructive impact on both the ecological and human worlds. The recent Black Summer bushfires in Australia, for example, burnt over 24 million hectares of land \cite{australia.royalcommissionintonationalnaturaldisasterarrangements_2020} and killed or displaced three billion animals \cite{lilymvaneeden.etal_2020}. At the same time, the bushfires directly caused 33 deaths \cite{australia.royalcommissionintonationalnaturaldisasterarrangements_2020}, reduced air quality to hazardous levels for humans \cite{divirgilio.etal_2021}, destroyed thousands of buildings and caused major disruption to critical infrastructure networks, bringing society to a standstill in affected areas \cite{filkov.etal_2020}. Climate scientists link the severity of the 2019/20 bushfire season---the most destructive ever recorded in Australia---with climate variability and long-term climate trends, which are themselves linked to technological changes since the Industrial Revolution \cite{abram.etal_2021, canadell.etal_2021, vanoldenborgh.etal_2021}. The Black Summer exemplified, in a brutal and disastrous manner, the consequences of changes in one world (technological) without considering linkages with other worlds (biophysical, ecological or human). 

Given the interdependence of the biophysical, ecological, human and technological worlds, participatory design researchers are increasingly concerned with the participation of entities beyond the human in the design of technology. Scholars of `posthuman design' and `more-than-human design', for example, are experimenting with design processes to shift design mindsets. Experiments include situating designers in the environment in which a technological artefact will operate \cite{bastian.etal_2018,wakkary_2021} and even encouraging designers to embody non-human entities in those environments, to consider the experience of other species or things \cite{clarke.etal_2019}. Yet, these experiments do not provide guidance on how to legitimate the perspectives of non-human entities through participatory design, instead focusing on situated or embodied experiences for designers to break down the boundaries between those designers and non-human entities.

In 2020, I worked with the Queanbeyan-Palerang Regional Council (`the Council'), as part of a team of researchers, to review the plans and progress of a smart city project in a region affected by the Black Summer bushfires. Our report recommended the Council adopt a more-than-human design framework, though we were unable to provide clear processes for the Council to enact the framework. This gap in design practice led to the review of theories of non-human behaviour from other disciplines, in particular philosophy and cognitive science, and eventually to the `intentional stance'.

To predict or explain the behaviour of other human or non-human entities, people often refer to the mental states of the other entity (their beliefs, desires or intentions) \cite{dennett_1987}. Dennett described this strategy of referring to the mental states of others to predict or explain their behaviour as ``adopting the intentional stance'' \cite{dennett_1987}. This paper explores how the intentional stance might be repurposed to legitimate the perspectives of non-human entities from the biophysical and ecological worlds, if those entities were to participate in the design process. Two questions guide the paper: 
\begin{itemize}
    \item How does adopting the intentional stance as a heuristic enable designers to account for the perspectives of non-human entities?
    \item What non-human entities could designers consider by adopting the intentional stance?
\end{itemize}

Each question is considered by reflecting on the smart city project. The paper builds on previous work with the Council, imagining how adopting the intentional stance might offer a process for the Council to enact the more-than-human design framework in the context of smart lighting systems. Through a process of experimentation and evaluation, the paper explores the potential of one approach (adopting the intentional stance) to clarify design processes for more-than-human participation.

\section{Background}
This section shares definitions of each of the `worlds', outlines posthuman or more-than-human design, and discusses the intentional stance.

\subsection{Biophysical, Ecological and Human Worlds}

The biophysical world is defined in this paper as the four spheres that support the existence of living things: the atmosphere (gases around the earth), the hydrosphere (water on earth), the lithosphere (rocks, soils and the earth's crust) and the biosphere (the sum of all ecosystems). The ecological world, on the other hand, is defined in this paper as living things on earth---plants and non-human animals---at the level of individuals, species, or individual ecosystems. Finally, the human world refers to individual humans and human societies.

The theme for the 2022 Participatory Design Conference (PDC) encourages authors and participants to explore how entities from worlds other than the human world may participate in design. This paper focuses on the connections between the biophysical, ecological, and human worlds, and explores how adopting the intentional stance could enable the participation of entities from the biophysical and ecological worlds in addition to the human world. 

\subsection{Posthuman Design and More-Than-Human Design}

Design has been dominated by a `human-centred' or `user-centred' paradigm since at least the 1980s \cite{forlano_2017,wakkary_2021}. These paradigms focus the attention of designers on the needs of individual human users and/or collectives of human users of technology, rather than technology itself. Recently, however, rapid transformation of the technological world alongside changes in the biophysical and ecological worlds have prompted a shift in the attention of designers towards the interrelations between human and non-human entities. An emerging posthuman design agenda encourages consideration of the non-human in the design process, including non-human animals and plants along with `things' and the artificial \cite{forlano_2017,giaccardi.redstrom2020Technology}. As Forlano \cite{forlano_2017} outlines, posthuman design has been influenced by theoretical approaches such as actor-network theory, feminist new materialism, object-oriented ontology, and transhumanism. Inheriting from these fields, the emerging posthuman design agenda encourages designers to transcend the limited focus of human-centred or user-centred design on human entities and consider a broader set of relations. Another term used to describe this emerging design philosophy is more-than-human design. Scholarly publications referring to more-than-human design have focused particularly on the design of smart cities \cite[\textit{e.g.},][]{clarke.etal_2019,loh.etal_2020,yigitcanlar.etal_2019}. More-than-human design scholars have also drawn on principles of relationality as understood in Indigenous epistemologies, which may enable designers to respectfully accommodate the non-human \cite{lewis.etal_2018,escobar_2018,akama.etal_2020}.

\subsubsection{Participation of Non-Human Entities}

For PDC 2022 we must ask -- what might be considered `participation' by non-human entities? Scholars of more-than-human design have experimented with two processes for non-human participation. In the first, designers are situated in the biophysical and ecological environment in which technology they design will operate, and encouraged to engage with non-human entities in that environment as part of the design process \cite{wakkary_2021,springgay.truman_2017, bastian.etal_2018}. This act is intended to break down barriers between the design process and the biophysical and/or ecological worlds in which designed technology will operate. Secondly, at PDC 2018 in Hasselt, Belgium, a group of designers undertook several experiments including embodying non-human animals as the designers navigated the city \cite{clarke.etal_2019}. This act was intended to encourage designers to ``see the city'' from the perspectives of non-human entities, which might then inform smart city design processes. However, as Akama et al. \cite{akama.etal_2020} note, participation implies the equivalent of ``human voice, rights, representation and structures of decision-making.'' While situated or embodied experiences for human designers may enable designers to `see' other worlds more clearly, there is a need to build on these experiments to design processes that legitimate the perspectives of non-human entities through participatory design, as we have for human entities.

\subsection{The Intentional Stance}

Dennett first proposed the intentional stance in 1971 \cite{dennett_1971} and included the idea in his book of the same name in 1987 \cite{dennett_1987}. Dennett argues that when we predict or explain the behaviour of an entity (humans, animals, artifacts---any object or system), we view that entity at varying levels of abstraction. In the first level, the physical stance, we predict the behaviour of an object or system based on its physical structure. We may abstract from the physical stance to the design stance, predicting the behaviour of a system based on our experience and knowledge of how the system is designed. Marchesi et al \cite{marchesi.etal_2019a} outline a useful example of adopting the design stance with respect to a car---we can predict that a car will slow down when we press the brake pedal based on our knowledge of the design of a car, rather than on our knowledge of the precise physical mechanisms that make up the braking system in a car. The most abstract of all the stances, the intentional stance, requires no knowledge of either physical structure or design. We predict the behaviour of another system from our {\itshape explanations} or {\itshape interpretations} of the beliefs and desires of the other system. Dennett \cite{dennett_1987} explains how we use this strategy as follows: 

\begin{quote}
{\itshape Here is how it works: first you decide to treat the object whose behavior is to be predicted as a rational agent; then you figure out what beliefs that agent ought to have, given its place in the world and its purpose. Then you figure out what desires it ought to have, on the same considerations, and finally you predict that this rational agent will act to further its goals in the light of its beliefs. A little practical reasoning from the chosen set of beliefs and desires will in many---but not in all---instances yield a decision about what the agent ought to do; that is what you predict the agent will do.} 
\end{quote}

Adopting the intentional stance might be a useful way to predict the behaviour of an entity if it were to participate in design activities, by treating the entity as if it had mental states. This does not lead to a theory of the operation of the internal mechanisms of any agent, and it does not require proof of the true intentionality of an agent. Rather it is a strategy for interpreting the behaviour of an entity {\itshape as if} it were a rational agent that governed its choice of action by considering its beliefs, desires or intentions---or mental states \cite{dennett_1987}. The strategy builds on the tendency of humans to attribute mental states to the entities whose behaviours we intend to interpret.  For this paper, then, there is a question as to the conditions under which an entity from either the biophysical or ecological world could be considered to have a mental state discernible to human designers. 

\section{Speculative Case Study: More-Than-Human Participation for Smart Lighting Systems}

In 2020, I worked as part of a team of researchers to review the plans and progress for the Queanbeyan Smart City Precinct ('smart city project'). Our report to the Council, overall, recommended an evolution in the design framework\textemdash from a techno-centred framework to a framework based on more-than-human design. While the overall recommendation was well received, the report lacked specific recommendations on how to enact the renewed framework. Our report encouraged the Council to think of the tensions between the interests of residents in a smart city and the interests of non-human entities in the ecological world, drawing on the lessons of the Black Summer bushfires. We also encouraged the Council to engage with Indigenous design practitioners during future design phases, to consider the relationality between smart city technologies and the people and places in which those technologies are embedded. Yet, we were unable to provide a clear process for Council staff themselves to consider the perspectives of entities within the region other than those of human residents, non-resident visitors or other humans managing the smart city. Without such a process, it is difficult to envisage the translation of the theories of posthuman or more-than-human design outside academic research contexts and into the everyday practice of smart city design. The lack of a clear design process to recommend to the Council led to the review of other disciplines, such as philosophy and cognitive science, for theories or heuristics that might translate into design processes. The section below presents an imagination of how adopting such a heuristic---the intentional stance---could legitimate the perspectives of two non-human entities towards the design of smart lighting systems in Queanbeyan. 

\subsection{Adopting the Intentional Stance to Legitimate the Perspectives of Trees and Air}

Many smart cities include smart lighting systems, including the smart city project administered by the Council. The lights in Queanbeyan shine brighter when pedestrians are in the vicinity of the lighting system, to create ``more vibrant areas with a feeling of safety'' \cite{queanbeyan-palerangsmartcitycouncil_}. In Queanbeyan, then, human and non-human entities must co-exist with artificial light at night, in variable intensities. The impact of artificial light on humans is well researched \cite[\textit{e.g.},][]{holker.etal_2010,cho.etal_2015}. Ecologists have also raised the alarm about the impact of artificial light on some non-human animals, such as sea turtles at the hatchling stage \cite{salmon_2003}. However, the impacts on plants and elements of the earth are largely unexplored. This section explores how adopting the intentional stance might allow designers to predict the behaviour of these entities, in the absence of definitive scientific guidance.

Plants possess chlorophyll which absorbs light to sustain their existence, but they also use light as a source of information \cite{bennie.etal_2016}. The natural cycles of light guide plants to regulate circadian rhythms and seasonal phenology, and also affect phenotypic variation (including variation in growth and resource allocation) \cite{bennie.etal_2016}. However, the widespread use of artificial lighting by human societies outdoors alters these cycles \cite{gaston.etal_2014}. Illumination from street lighting is brighter than and inconsistent with natural cycles of moonlight, resulting in many plants experiencing artificial light at night at levels consistent with physiological effects (even if in short duration) \cite{bennie.etal_2016}.

If we adopt the intentional stance towards trees located in an area proposed for a smart lighting system (entities from the ecological world), we would treat those trees as rational agents. Based on the points raised in the paragraph above, one might say that trees would believe in the importance of growth and resource allocation consistent with natural daily and seasonal cycles of light. The trees' desires and intentions, then, could include absorbing light within the optimal spectrum in natural cycles to optimise photosynthesis. Following the process outlined by Dennett for the intentional stance, a designer would predict that a tree ought to seek out light within that spectrum and during those cycles. By extension, trees located within an area proposed for a smart lighting system, participating in the design of the smart city, would advocate for minimal disruption to lighting within those parameters.

We might also wish to consider the perspectives of particles of air surrounding a smart lighting system\textemdash entities from the biophysical world. Excess artificial light at night can contribute to air pollution by interfering with chemical reactions that naturally clean the air during the night \cite{stark.etal_2010}. Chemicals emitted from a variety of human activities are broken down by a form of nitrogen oxide called the nitrate radical, which only occurs in darkness as sunlight destroys the naturally occurring chemical \cite{brown.stutz_2012}. We might use this information to try and predict how particles of air would act to avoid such disruption to air filtration. Yet it is difficult to discern the mental states of particles of air that would motivate such behaviour---what does a particle of air believe, desire or intend to do, other than exist? Adopting the intentional stance first requires identifying such mental states, which may not be possible for non-living entities. In addition, it is difficult to identify individual, localised entities in the atmosphere---could the intentions of individual air particles be determined separately from the air as a whole? If not, how can we identify entities from the atmosphere that should participate in the design of a particular smart lighting system or smart city?

\section{Discussion}

This section discusses the usefulness and limitations of the intentional stance as a heuristic for more-than-human participatory design, and briefly presents alternatives for consideration.

\subsection{The Intentional Stance as a Heuristic}

For the ecological world, this paper demonstrates through experimentation with the example of a smart lighting system that the intentional stance could become a heuristic for designers to predict the behaviour of some non-human entities, such as non-human animals and plants. Drawing on the case study, the Council could consider the perspectives of trees alongside the perspectives of human stakeholders in the smart city\textemdash residents, non-resident visitors, and those involved in the management of the smart city. The goal of such a process is not to exclude the perspectives of human stakeholders, but to balance the perspectives of humans alongside those of entities from other worlds. For example, residents and non-resident visitors may desire that the pavement in a certain area is illuminated alongside tree cover to improve accessibility and promote safety. Following the process enabled by adopting the intentional stance, designers may then consider tensions between the interests of human stakeholders and those of trees, as outlined earlier. On the other hand, interpreting the mental states of an element that makes up part of the biophysical world\textemdash the air\textemdash was not straightforward. For the biophysical world, then, adopting the intentional stance loses value as a simple heuristic for designers (or anyone practicing design) to predict the behaviour of non-human entities.

Entities from different worlds may require different methods or processes to enable their legitimate participation in technology design. However, we must not wait for an all-encompassing approach to participatory design for any non-human entity. The design of technological systems continues apace, guided by techno- or human-centred design philosophy, and heuristics may assist the practice of everyday design to contribute, now, to avoiding ecocidal futures. Instead, I encourage application, critique and re-imagination of the intentional stance for entities from the ecological world, just as this paper was developed on the foundation set by earlier experiments in more-than-human design.

\subsection{Limitations of the Intentional Stance}

There are several limitations to adopting the intentional stance for the participation of non-human entities from the ecological world, notwithstanding its potential as a heuristic. Firstly, non-human entities may not act within the rational dynamic assumed by the intentional stance. In the case of humans, social psychology has documented our irrationality for decades \cite{tversky.kahneman_1974}. More recently, affective neuroscience has demonstrated the centrality and importance of emotions to human decision making \cite{mlodinow_2022}. Further research might consider how non-human entities, from either the biophysical or ecological worlds, act outside the bounds of rationality, and how that behaviour may be incorporated into or considered separately from the process prescribed by adopting the intentional stance.

Secondly, the intentional stance is focused on predicting the behaviour of an individual agent or aggregates of individual agents, but not collectives of agents. As shown in this paper, we may adopt the intentional stance to discern the mental states and predict the behaviour of a tree, but what about a forest? How might the perspective of a forest conflict with the perspective of an individual tree? What additional processes are required, or how can the intentional stance be amended, to consider the perspectives of collectives?

Thirdly, adopting the intentional stance for more-than-human participatory design assumes that humans can and should `speak for' non-human entities. The perspective of a non-human entity predicted by a designer following such a process will inevitably be imbued with the perspective of the designer towards the non-human entity. For example, the information presented in the case study on the physiology of trees, and how physiology might be affected by artificial light, relied solely on desktop research conducted for this study. The perspective of the tree was informed by an assessment of the validity and reliability of evidence available in academic research publications. Further research may add steps to the process to guide reflexive practice by designers adopting the intentional stance\textemdash identifying sources of information and interrogating positionality\textemdash to enable outsiders to contextualise human predictions of non-human behaviour.

Finally, the case study adopts the intentional stance to interpret the perspectives of a limited set of entities---entities within the immediate vicinity of a smart lighting system. Further research could extend the boundary of analysis to consider how such a design process might include consideration of the impact of a smart lighting system more broadly. For example, one could extend the analysis to consider the perspectives of entities impacted by the supply chain for the smart lighting system (\textit{e.g.}, the factory producing sensors that manage the intensity of light and the source of energy for the Council area).

\subsection{Alternatives to the Intentional Stance}

\subsubsection{Direct Sensing As Participation}
For a more direct approach to more-than-human participatory design than adopting the intentional stance, we could consider using environmental sensors to capture signals from ecological or biophysical worlds to inform design processes, or to convert signals from sensors into inputs for the operation of a technological system. One might argue that using trees, for example, as networked elements of a technological system reflects a techno-centred design philosophy. However, identifying a `goal state' for a tree (\textit{e.g.}, defining indicators of a thriving tree), assessing the physiological health of trees in Queanbeyan against those metrics over time, and adjusting the operation of technology in response to signals might be one way to procure `participation' of non-human entities.

\subsubsection{Engaging With Indigenous Epistemologies}
The growing engagement with Indigenous epistemologies in design research is a welcome expansion of cultural perspectives in our research community. As argued by Akama et al.\cite{akama.etal_2020}, there is much for more-than-human design to learn from cultural perspectives that prioritise pluriversal and relational thinking and being. The quest for simple heuristics to guide everyday designers to enact a more-than-human design framework is not intended to limit or distract from the embrace of Indigenous epistemologies. Rather, it is intended as a compliment to those efforts and is, indeed, informed itself by the relational perspectives of those epistemologies. In addition, I encourage critique of the heuristic proposed in this paper from Indigenous perspectives.

\section{Conclusion}

Posthuman and more-than-human design scholars are shifting the attention of the design community from the prevailing techno- or human-centred design paradigms towards the interrelations between humans and non-human entities. This shift is urgent\textemdash the biophysical and ecological worlds that sustain us physically and spiritually are rapidly transforming due, at least in part, to our insensitivity to the interdependencies of life on earth. To translate renewed design philosophies from the realms of academic research contexts into the everyday practice of design, we need to develop clear processes to enact more-than-human design. This paper explored the usefulness of adopting the intentional stance\textemdash a strategy for interpreting the behaviour of non-human entities drawn from philosophy\textemdash as a heuristic for more-than-human participatory design. I encourage application, critique, re-imagination and, even, formalisation of the intentional stance to enable the participation of non-human animals and plants from the ecological world, considering the limitations raised in this paper. Further research on methods of participation for entities from the biophysical world is also encouraged, along with consideration of alternative approaches to legitimate the perspectives of entities from the ecological world. Without developing such heuristics for everyday design, visions of more-than-human design for smart cities proffered by our research community risk being sidelined in the practice of technology design.

\begin{acks}
My sincere thanks to Alex Zafiroglu, Elizabeth Williams, Ben Swift, Josh Andres, Danny Bettay and anonymous reviewers for thoughtful feedback and comments. Thank you also to members of the Queanbeyan-Palerang Regional Council and my colleagues from the Australian National University---Lorenn Ruster, Nischal Mainali and Teffera Teffera---who engaged in the project that preceded this paper.
\end{acks}

\bibliographystyle{ACM-Reference-Format}
\bibliography{main.bib}

\end{document}